\documentclass[aps,prl,twocolumn,showpacs,superscriptaddress,groupedaddress]{revtex4}
\usepackage{natbib}
\usepackage{amsmath,amssymb,amsfonts,subfigure,braket,color,dsfont}
\usepackage[english]{babel}
\usepackage{graphicx} 
\usepackage{sistyle}
\usepackage{listings}
\usepackage{gensymb} 

\begin{document} 
 
\title{Double-heterodyne probing for ultra-stable laser based on spectral hole burning in a rare-earth doped crystal }
\author{N. Galland} 
\affiliation{Univ. Grenoble Alpes, CNRS, Grenoble INP and Institut N\' eel, 38000 Grenoble, France}
\affiliation{LNE-SYRTE, Observatoire de Paris, Universit\' e PSL, CNRS, Sorbonne Universit\' e, Paris, France}
\author{N. Lu{\v c}i\'c} 
\affiliation{LNE-SYRTE, Observatoire de Paris, Universit\' e PSL, CNRS, Sorbonne Universit\' e, Paris, France}
\author{S. Zhang} 
\affiliation{LNE-SYRTE, Observatoire de Paris, Universit\' e PSL, CNRS, Sorbonne Universit\' e, Paris, France}
\author{H. Alvarez-Martinez} 
\affiliation{LNE-SYRTE, Observatoire de Paris, Universit\' e PSL, CNRS, Sorbonne Universit\' e, Paris, France}
\affiliation{Real Instituto y Observatorio de la Armada, San Fernando, Spain}
\author{R. Le Targat} 
\affiliation{LNE-SYRTE, Observatoire de Paris, Universit\' e PSL, CNRS, Sorbonne Universit\' e, Paris, France}
\author{A. Ferrier} 
\affiliation{Chimie ParisTech, Universit\' e PSL, CNRS, Institut de Recherche de Chimie Paris, 75005 Paris, France} 
\affiliation{Sorbonne Universit\'e, Facult\'e des Sciences et Ing\'enierie, UFR 933, 75005 Paris, France} 
\author{P. Goldner} 
\affiliation{Chimie ParisTech, Universit\' e PSL, CNRS, Institut de Recherche de Chimie Paris, 75005 Paris, France} 
\author{B. Fang} 
\affiliation{LNE-SYRTE, Observatoire de Paris, Universit\' e PSL, CNRS, Sorbonne Universit\' e, Paris, France}
\author{S. Seidelin}
\affiliation{Univ. Grenoble Alpes, CNRS, Grenoble INP and Institut N\' eel, 38000 Grenoble, France}
\affiliation{Institut Universitaire de France, 103 Boulevard Saint-Michel, F-75005 Paris, France}
\author{Y. Le Coq}
\affiliation{LNE-SYRTE, Observatoire de Paris, Universit\' e PSL, CNRS, Sorbonne Universit\' e, Paris, France}

\date{\today}

\begin{abstract}
We present an experimental technique for realizing a specific absorption spectral pattern in a rare-earth-doped crystal at cryogenic temperatures. This pattern is subsequently probed on two spectral channels simultaneously, thereby producing an error signal allowing frequency locking of a laser on the said spectral pattern. Appropriate combination of the two channels leads to a substantial reduction of the detection noise, paving the way to realizing an ultra-stable laser for which the detection noise can be made arbitrarily low when using multiple channels. We use such technique to realize a laser with a frequency instability of $\mathbf{1.7\times10^{-15}}$ at 1 second, not limited by the detection noise but by environmental perturbation of the crystal. This is comparable with the lowest instability demonstrated at 1 second to date for rare-earth doped crystal stabilized lasers.   
\end{abstract}


\maketitle

Ultra-stable continuous wave (cw) lasers are essential tools in many high precision measurements, including in particular optical clocks \cite{Ushijima_2015,Nicholson_2015,Tyumenev_2016,Schippo_2017}. Stabilizing a laser on a spectral pattern photo-imprinted by spectral hole burning (SHB) in a rare-earth-doped material at cryogenic temperature \cite{Chen_2011,Cook_2015,Gobron_2017} is a promising technique. Compared to high finesse Fabry-Perot (FP) cavity stabilized lasers, where considerable efforts are currently dedicated to reducing the fundamental thermodynamic noise, by increasing the cavity mode volume \cite{Falke_2015}, using crystalline materials for the spacer and mirror \cite{Matei_2017} or the mirror coatings \cite{Cole_2013}, cooling down to cryogenic temperatures \cite{Robinson_2019}, or a combination of these approaches, SHB based lasers are estimated to have a much lower thermodynamic limit, and may therefore equal or surpass the achievable performance of cavity stabilized lasers.

In our previous work \cite{Gobron_2017}, we described an apparatus demonstrating frequency locking of a cw laser onto a narrow spectral hole using a crystal of yttrium ortho-silicate doped with 0.1\% europium atoms, where the $2\times10^{-14}$ fractional frequency instability at 1\,s was limited by the detection noise. In this work we demonstrate novel double-heterodyne probing methods with reduced detection noise. We utilize it to produce a laser with $1.7\times10^{-15}$ fractional frequency instability at 1\,s (as measured via a frequency-comb), limited by environmental perturbation of the crystal rather than the detection noise.


Our experimental apparatus has been described previously \cite{Gobron_2017}. Note however we now use a different closed-cycle cryostat (customized OptiDry by MyCryoFirm) which, contrary to the previously used system, does not require the addition of an extra passive isolation stage in the cryostat to obtain acceptably low vibration levels for our application. The device will be the subject of a separate publication.

\begin{figure}[htbp]
	\centering
	\includegraphics[width=\linewidth]{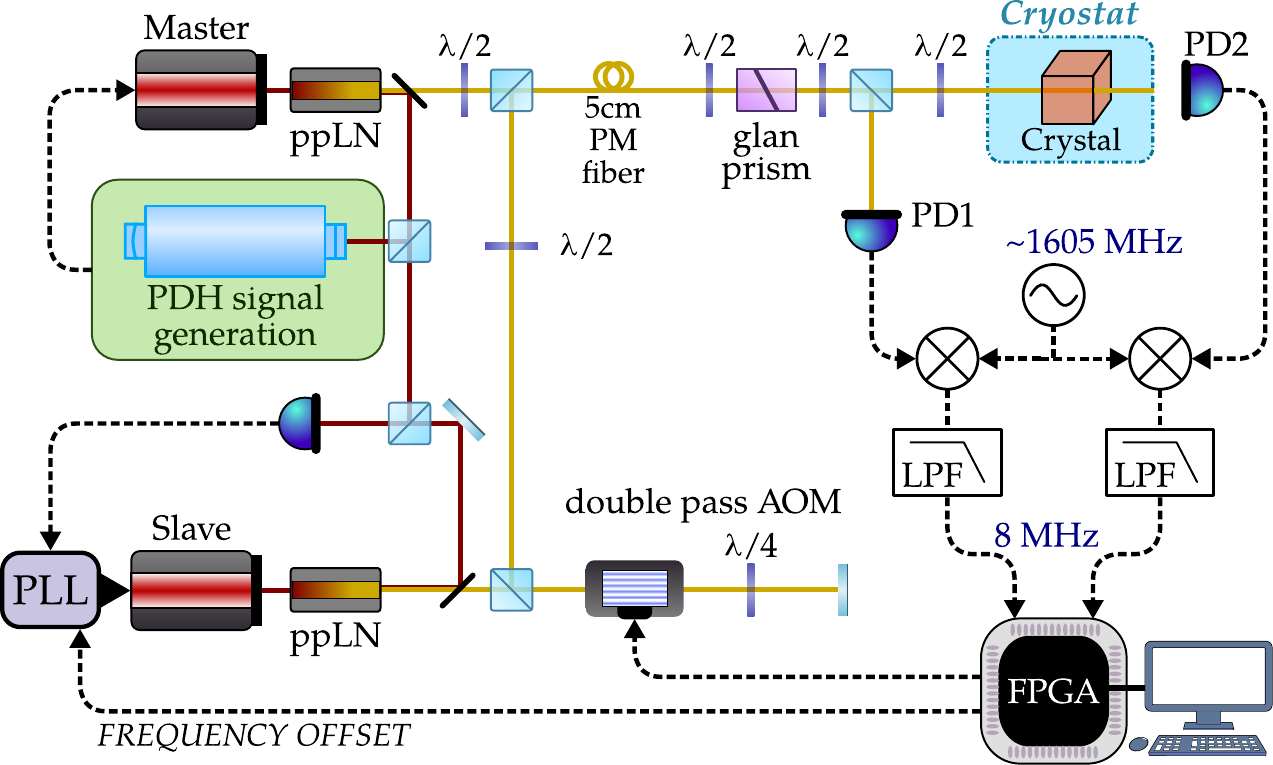}
	\caption{Schematics of the experimental setup for creating and interrogating spectral holes. ppLN: periodically poled lithium niobate. PDH: Pound-Drever-Hall. PM fiber: polarization maintaining fiber. PD: photodiode. PLL: phase lock loop. AOM: acousto-optic modulator. LPF: low-pass filter. FPGA: field programmable gate array.}
	\label{fig:setup}
\end{figure}

Briefly speaking, the optical setup, presented in figure \ref{fig:setup}, is based on two commercial extended cavity diode lasers, referred as master and slave in the following, and emitting 65\,mW at 1160\,nm. Both lasers are fiber coupled and frequency doubled through ppLN waveguides to 580\,nm output in free space. We obtain 8\,mW of master and 5\,mW of slave at 580\,nm, corresponding to the absorption wavelength of the ${^7}F_0\longrightarrow{^5}D_0$ transition in $\rm Eu^{3+}{:}Y_{2}SiO_{5}$. We use the absorption peak at 580.039 vacuum wavelength (site 1) throughout this work.

As the doubling process is not perfect, a part of the infrared light is transmitted by the ppLN waveguides, can be separated from the 580\,nm light using dichroic mirrors, and is used for frequency/phase locking of the lasers. The master laser at 1160\,nm is frequency locked to a 10\,cm long commercial ultra stable cavity using the PDH technique. The slave laser is phase locked to the master laser with a tunable frequency offset that can be changed in real time by a digital control system based on a software defined radio (SDR) platform. A part (1\,mW) of the 1160\,nm light from the slave laser is also sent (via a noise canceled optical fiber) to an optical frequency comb, located in another room, for characterization. With this setup, the slave laser benefits from the stability and linewidth of the master laser, with a continuously tunable frequency offset between the two. This allows to set the slave laser frequency at the maximum of the europium ions inhomogeneously broadened absorption while the master laser frequency is maintained far from it. It also provides a feedback mechanism to frequency locking the slave laser on a spectral pattern realized in the rare-earth doped crystal.

The slave laser light at 580\,nm is passing through a double pass AOM which allows to control both the optical power of the beam and the spectral properties of the light in a $\pm$1\,MHz range around the laser frequency. Appropriate use of our digital control system allows generation of arbitrary spectral light pattern with this AOM. The slave light is then combined with the master light at 580\,nm, matched in spatial mode, and polarized along the D$_1$ crystal axis for optimal absorption \cite{Ferrier_2016}. Half of the light goes through the rare-earth doped crystal located in the cryostat and is collected in a photodiode (EOT-2030A). The other half is directly sent to an other identical photodiode. Appropriate comparison of the spectral properties of the signals obtained on the two photodiodes provide direct access to both the absorption and dispersion properties of the $\rm Eu^{3+}{:}Y_{2}SiO_{5}$ crystal.

The beatnote obtained on each photodiode (around 1.6\,GHz) is then demodulated to 8\,MHz (using triple balanced mixers and a common mode synthesizer whose frequency takes automatically into account the frequency difference between master and slave lasers) and amplified before digitization in each input of the SDR platform. This same SDR platform is used to drive the double pass AOM and controls the offset frequency between master and slave.

The crystal is a $8\times8\times4\mathrm{~mm^3}$ parallelepipoid of europium doped yttrium orthosilicate ($\rm Eu^{3+}{:}Y_2SiO_5$) grown by Czochralski method with 0.1\% europium doping. The optical beam propagates along the crystallographic \emph{b} axis and the two largest facets, perpendicular to this axis, are polished. The crystal temperature is maintained and stabilized around 3.7 Kelvin in a commercial pulse tube cryostat. At such low temperatures, long-lived spectral features can be photo-imprinted in the crystal by the SHB technique. Applying a laser radiation of a fixed optical frequency within the inhomogenously broadened absorption spectrum of the crystal, at a sufficiently high power and long enough duration (we typically use $65\,\mu\mathrm{W\cdot cm^{-2}}$ and 0.5\,s), the class of $\rm Eu^{3+}$ doping ions resonant with the radiation will be selectively depopulated from their initial hyperfine state and will accumulate in other hyperfine states, which are metastable at sufficiently low temperatures (lifetime of several hours at 3.7\,K) and dark-states for the optically pumping radiation. 


In a previous work, we used such apparatus to probe dispersively a single narrow spectral hole, thereby extracting an error signal providing the necessary information to frequency lock the probe laser to the spectral hole. This error signal typically exhibits a 0.18\,mrad/Hz discriminator slope when probing a 4\,kHz broad spectral holes with 40 percent contrast. This resulted in a laser with $2\times10^{-14}$ fractional frequency stability at 1\,s, where we identified the stability to be limited by the noise of the probing method. The optical power used for the probe laser that interacts with the narrow spectral hole was then set to 100\,nW ({\it i.e.} $0.26\,\mu\mathrm{W\cdot cm^{-2}}$), which was experimentally found to be suitable for long-time (couple of hours) operation. Note that here, a compromise must be made between high signal (inducing higher signal-to-noise ratio in the detection) and long-term operation, as the probing optical power itself is slowly overburning the hole, leading to hole degradation (decrease of the discriminator slope) over time.

\begin{figure}[t]
	\centering
	\includegraphics[width=\linewidth]{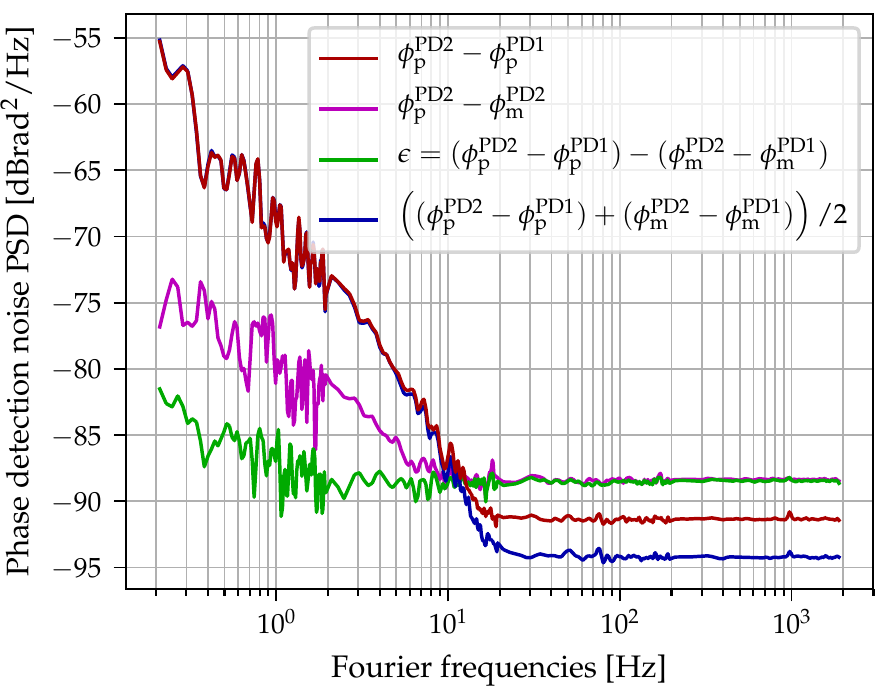}
	\caption{Power spectral density of background phase noise for various detection configurations. The phase noise of $\phi_{\rm p}^{\rm PD2}$, not represented, is largely (several tens of dB) above all the curves shown.}
	\label{fig:detectionPSD}
\end{figure}

Even after improvement of the radio frequency (RF) chain with low noise components at critical parts, by measuring the intrinsic noise of the probing method (red curve in fig.~\ref{fig:detectionPSD}), we observe a white excess phase noise plateau for high Fourier frequencies, with a steep rise at longer timescales (lower Fourier frequencies), which produces excess frequency fluctuations when this phase measurement is used to derive an error signal for frequency locking the laser on the spectral hole. This excess phase noise at low Fourier frequency is therefore a major feature that requires improvement.

\begin{figure}
	\centering
	\includegraphics[scale=1]{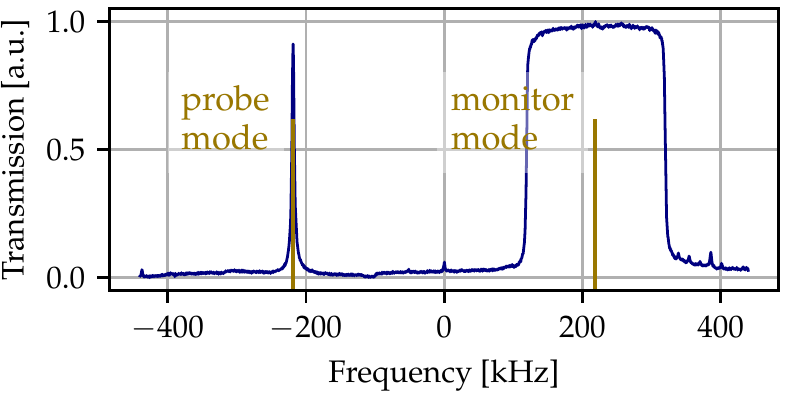}
	\caption{Example of a pattern in the absorption spectrum of the crystal corresponding to a double-hole structure, with one narrow and one broad (200\,kHz) spectral hole.}
	\label{fig:doubleholes}
\end{figure}

We have observed this phase noise to be a combination of additive and multiplicative noises appearing in the laser propagation, photodetection, amplifications, demodulations and digitizing chains. In particular, by increasing the optical power of the slave laser in both detection arms, we have been able to proportionally lower the white phase noise plateau at high Fourier frequencies (characteristic of an additive noise), whereas the noise level at lower Fourier frequency remains unchanged (characterizing a multiplicative noise, that could arise, for example, from parasitic FP cavities or path length fluctuations). In order to decrease the multiplicative noise, we have explored using a supplementary beatnote signal, relatively close in frequency to the one used to derive the dispersive error signal, to monitor and largely suppress the effect of such noise sources. In practice, this is realized by burning not only a narrow spectral feature on which the laser is frequency locked, but also a broad flat hole next to it (see fig.~\ref{fig:doubleholes}), used to transmit a supplementary optical mode (so-called monitor mode) that produces the extra beatnote with the master laser. To create such a structure, the frequency of the slave laser is swept over a few hundreds of kilohertz while applying the same power as the one used to create the narrow hole for a few tens of seconds (scanning speed being of the order of 20\,kHz/s). During probing, appropriate RF wave is applied to the AOM, so as to generate a slave laser radiation composed of 2 cw modes (in the spirit of \cite{Jobez_2016}), one (probe mode) close to resonance with the narrow spectral hole, and the other (monitor mode) close to the center of the broad flat spectral hole. These two modes have a fixed optical frequency difference (typically 450\,kHz) and phase relationship. In this spectral feature configuration, we utilize the fully digital nature of our measurement system to derive a modified error signal from the phase information of the various components beating on PD1 and PD2.

Data processing in the Fourier space by the SDR platform, using the Fast Fourier Transform algorithm, allows extracting the phases of the beatnotes (in PD1 and PD2 signals) between the master laser and the probe mode $\phi_{\rm p}^{\rm PD1,PD2}$, as well as those between the master laser and the monitor mode $\phi_{\rm m}^{\rm PD1,PD2}$. From these, an error signal proportional to $\epsilon = (\phi_{\rm p}^{\rm PD2}-\phi_{\rm p}^{\rm PD1})-(\phi_{\rm m}^{\rm PD2}-\phi_{\rm m}^{\rm PD1})$ is digitally constructed. At first approximation, only $\phi_{\rm p}^{\rm PD2}$ is impacted by the offset between the center of the narrow spectral hole and the slave laser (the monitor mode of the slave laser, propagating in a broad spectral hole, sees negligible dephasing when a frequency offset appears). As a consequence, this error signal is appropriate for use in an active feedback loop aiming at locking the slave laser onto the narrow spectral hole. However, the other components of $\epsilon$ allows rejection of some detection noise sources. High Fourier frequencies of the detection noise are characterized by a white noise plateau of additive noise, the amplitude of which is well accounted for by the dark noise of the photodiodes in use. On the contrary, low Fourier frequencies exhibit excess noise which is largely removed by using the linear combination defining $\epsilon$. More precisely, $\phi_{\rm p}^{\rm PD2}$ in itself is strongly impacted by technical noise, making it unsuitable for practical servo realization. However, as can be seen in figure \ref{fig:detectionPSD}, a large part of these phase fluctuations are removed by subtracting from the measured $\phi_{\rm p}^{\rm PD2}$ either the technical phase fluctuations monitored by the same mode on PD1 ($\phi_{\rm p}^{\rm PD1}$), or the technical phase fluctuations as monitored by the second propagating optical mode ($\phi_{\rm m}^{\rm PD2}$). Combining these two approaches in $\epsilon$ makes detection noise negligible compared to the currently achieved stability of the laser locked to spectral hole (see below).

The price to pay for using this largely immune to excess noise combination $\epsilon$ is a 3\,dB increase of the phase noise at high Fourier frequencies, stemming from the double impact of additive noise in a photodiode when combining the respective phases of two RF signals (whereas, when combining the phases of RF signals detected on two different photodiodes, these additive noises which are uncorrelated are thus summed quadratically).

It should be emphasized, however, that the limitation due to additive noise is largely dependent on the optical power used to probe the spectral hole (a fact we verified experimentally over a large range of powers), the presented value of 100\,nW per mode being a reasonable compromise showing negligible overburning for a couple of hours of continuous operation. Furthermore, by probing several narrow holes in parallel, it will be possible to reduce this effect proportionally to the square root of the number of holes. As a proof-of-principle, we have exemplified this last effect in fig. \ref{fig:detectionPSD}, on the curve presenting the $\left( (\phi_{\rm p}^{\rm PD2} - \phi_{\rm p}^{\rm PD1}) + (\phi_{\rm m}^{\rm PD2} - \phi_{\rm m}^{\rm PD1}) \right)/2$ phase noise, which can be interpreted as using the monitor mode as if it were probing a second narrow spectral hole. In this case, the additive noise is reduced by 3\,dB compared to the simple heterodyne detection. Note that in this last construct, the low Fourier frequency noise is unchanged compared to the simple heterodyne detection, further exemplifying its nature as a mostly common mode to the different channels (multiplicative noise), which explains why it is largely suppressed when using $\epsilon$ as an error signal.  


\begin{figure}[t]
	\centering
	\includegraphics[width=\linewidth]{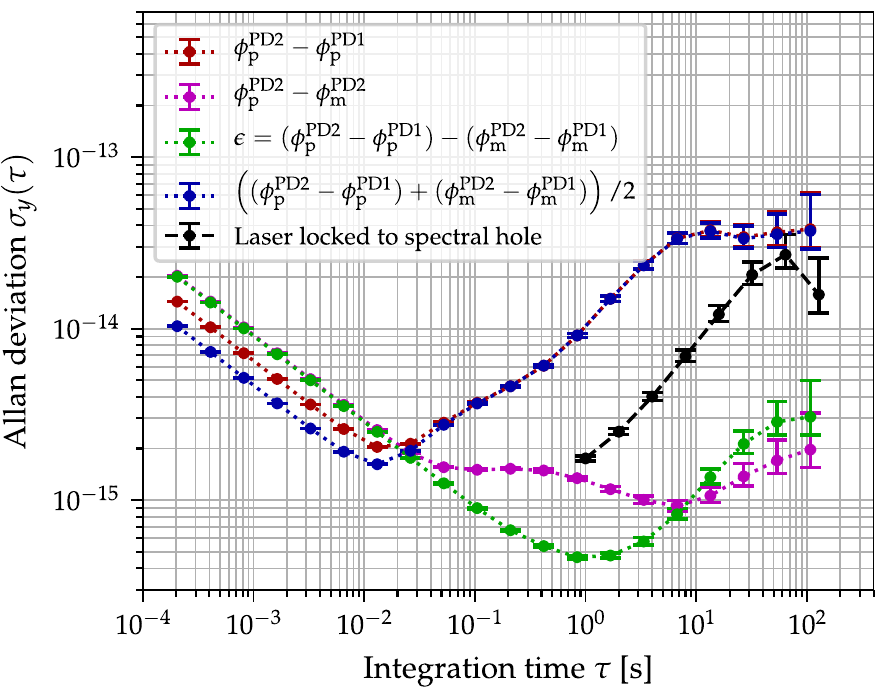}
	\caption{Fractional frequency instabilities (Allan deviations). Black line: laser locked by double-heterodyne probing method to a spectral hole (measured via optical frequency comb). Colored lines, projected instability stemming from measured detection noise (for a discriminator slope of 0.18\,mrad/Hz) in various configurations.}
	\label{fig:laserstability}
\end{figure}

Using this double heterodyne probing technique based on the $\epsilon$ error signal construct, we have significantly improved the stability of a laser frequency locked to a hole, as shown in figure \ref{fig:laserstability}. The presented stability curve is limited by environmental stability and not by the detection noise. In fact, the main limitations can be attributed, with high certainty, to a combination of temperature fluctuations of the crystal and uncompensated optical path fluctuations between the crystal and probing beam, as no special efforts have been made here yet to mitigate these effects, contrary to ref. \cite{Cook_2015}. Our double-heterodyne detection scheme therefore proves useful to reach the current technical limitation of the experimental apparatus. Furthermore, it is a proof-of-principle demonstration of how appropriately combining signals measured on different channels in the vicinity of specifically engineered spectral hole pattern can be used both for lowering detection noise at high Fourier frequencies (additive noise) and at low Fourier frequencies (largely common mode for the different channels). Even though we only use 2 modes at the moment, in the future, multiplying probe modes (for lower additive noise) and monitor modes (for better rejection of multiplicative noise) will prove a strong benefit for the combination of spectral hole pattern engineering and digital extraction of an appropriate error signal. Comparing to the more traditional technique based on the PDH method generation of error signal with analog electro-optic modulator (EOM)-based phase-modulation of the probe mode and analog demodulation after traversing the dispersive crystal, as utilized in \cite{Cook_2015}, we believe the presented proof-of-principle scheme present some advantages. The power in the side modes (modulation peaks) in the PDH technique is limited by the EOM operating principle, thereby reducing the achievable signal-to-noise ratio for a given probe power. In our scheme, we can increase the power in the monitor modes much above the maximum achievable in an EOM, within the practical limits set by, in particular overburning from the monitor modes. Furthermore, the scheme we present in this letter is, in large part, immune to some class of experimental noise classically impacting PDH detection, in particular, parasitic FP cavities: the monitor mode is in fact probing such parasitic effect and removing it in the $\epsilon$ construction (except for the small mismatch between the optical frequencies of the probe and monitor modes). Future extensions of our scheme involving multiple monitor modes (allowing better estimation of the exact parasitic FP etalon effect at the probe optical frequency) may allow even higher rejection of such effects. Such possibility does not seem possible to implement in the classic PDH scheme, which is, by design, sensitive to FP etalon effects. Likewise, a traditional issue of PDH implementations is the detrimental effect of residual amplitude modulation (RAM) in the sideband generating EOM \cite{Wong_1985}. We operate here in a totally different regime in which we recover the phase information digitally, thereby {\it a priori} strongly alleviating RAM-like effects. A detailed study of RAM effects goes beyond the scope of this paper but will certainly be interesting to explore in the future, the possible role of amplitude noise to phase noise conversion in photodetectors being an open question.


Exploring the possibilities of using complex data treatment protocols, rendered accessible by nowadays digital platforms, is an important topic that may lead to detection schemes exhibiting both high signal-to-noise ratio and substantial immunity to a large class of technical noise sources. As a proof-of-principle, we have demonstrated here such a scheme, lowering the detection noise induced instability at 1\,s of a spectral hole stabilized laser in the low $10^{-16}$ range at 1\,s, where environmental stability of the crystal is limiting the frequency stability of the spectral holes (and not the detection noise). We have also demonstrated the possibility of further reducing the detection noise with parallel probing of multiple narrow holes, at Fourier frequencies where additive noise is the dominating limitation. Future work will involve, of course, improvement of the stability of the environment of the crystal (temperature stability, vibration reduction, electromagnetic shielding, noise cancelling of critical optical paths...) but also further lowering the detection noise (both additive and multiplicative) by appropriately multiplying the number of probe and monitor modes. Such studies pave the way towards realizing ultra high stability lasers with unprecedented performance, which will find direct application in particular in optical lattice clocks to help them achieve the quantum projection noise limit. Another direct application of the ultra high precision measurement capability provided by our techniques is related to realizing quantum sensors, such as accelerometers (as mentioned in ref. \cite{Galland_2019} for example) or micro-fabricated force-sensors in the spirit of ref. \cite{Molmer_2016,Seidelin_2019}. Finally, exploring the use of such multi-heterodyne probing for other physical system-stabilized lasers (FP cavity, atomic or molecular reference,...) may also be an interesting extension of such work.  

This project has received funding Ville de Paris Emergence Program, LABEX Cluster of Excellence FIRST-TF (ANR-10-LABX-48-01), within the Program ``Investissements d’Avenir'' operated by the French National Research Agency (ANR), and European Union’s Horizon 2020 research and innovation program under grant agreement No 712721 (NanOQTech).

\bibliography{sample}

\end{document}